# Interdisciplinarity at the Journal and Specialty Level: The changing knowledge bases of the journal *Cognitive Science*,

Loet Leydesdorff*[a] & Robert L. Goldstone[b]

**Abstract**

Using the referencing patterns in articles in *Cognitive Science* over three decades, we analyze the knowledge base of this literature in terms of its changing disciplinary composition. Three periods are distinguished: (1) construction of the interdisciplinary space in the 1980s; (2) development of an interdisciplinary orientation in the 1990s; (3) reintegration into "cognitive psychology" in the 2000s. The fluidity and fuzziness of the interdisciplinary delineations in the different visualizations can be reduced and clarified using factor analysis. We also explore newly available routines ("CorText") to analyze this development in terms of "tubes" using an alluvial map, and compare the results with an animation (using "visone"). The historical specificity of this development can be compared with the development of "artificial intelligence" into an integrated specialty during this same period. "Interdisciplinarity" should be defined differently at the level of journals and of specialties.

[a] University of Amsterdam, Amsterdam School of Communication Research (ASCoR), Kloveniersburgwal 48, 1012 CX Amsterdam, The Netherlands; loet@leydesdorff.net ; * corresponding author.
[b] Department of Psychological and Brain Sciences, Indiana University, USA; rgoldstone@indiana.edu.

## Introduction

Did "cognitive science" emerge as an interdisciplinary field among psychology, linguistics, computer science, philosophy, and (increasingly) the neurosciences during the last few decades? (Gentner, 2010; Goldstone & Leydesdorff, 2006; Leydesdorff, Goldstone, & Schank, 2008; Shunn *et al.*, 1998; Van den Besselaar, in preparation; Von Eckhardt, 2001). The journal *Cognitive Science* published its first volume in 1977, and became established as the journal of the Cognitive Science Society with the publication of its fourth volume in 1980. On this occasion, the Editor of the journal stated its mission as follows:

> "Cognitive Science is multidisciplinary, requiring tools and insights from many different scientific areas. We intend to broaden the range of articles published in the journal to include more linguistics, philosophy, developmental psychology, cognitive anthropology, the neurosciences, etc. The goal is that the Society and journal should reflect the broad range of interests and knowledge required for the emergence of a substantive science of cognition" (Collins, 1980, at p. i).

Despite this ambition to reach out to other disciplines, the journal has remained most influenced by psychology, and, in fact, has arguably become lodged within psychology increasingly over the years. As Gentner (2010) noted recently: "The proportion of papers authored by psychologists has increased steadily from 1978, when psychologists constituted about a quarter of the authors, to 2008 when psychologists constituted over half of the contributors. If the proportion doubles again in the next 30 years, by 2038 we will have vanquished the other fields entirely and established total dominion over Cognitive Science. But such a coup would be a Pyrrhic victory" (p. 330).

The attempt to make the journal interdisciplinary has sometimes been an uphill battle because new disciplinary structures may evolve over time. Cognitive scientists generally see the value of interdisciplinarity in fostering cross-fertilization among areas, tackling applied problems that require expertise that falls outside of any single area, and providing multiple, mutually-illuminating perspectives on questions of common interest. However, "interdisciplinarity" may not provide a stable equilibrium.

In this study, we address the question of interdisciplinarity by focusing on the history of this journal using the instruments of bibliometric and network analysis. The research question emerged first as a follow-up question in the study of the emergence of artificial intelligence by Van den Besselaar & Leydesdorff (1996, at p. 428). These authors concluded that it was not possible at the time to distinguish a set of journals as "cognitive science" and suggested using the "interfactorial complexity" of this core journal of the field as an indicator of interdisciplinarity. Van den Besselaar & Heimeriks (2001) further developed this indicator, whereas Goldstone and Leydesdorff (2006)—one of us a former Editor of the journal—further analyzed the citation and referencing patterns of this journal and found an imbalance between the import and export of knowledge when operationalized in terms of citation rates.

In 2004, articles appearing in *Cognitive Science* drew mainly from psychology and its neighboring disciplines for their cited references, but were cited by a much wider arena, including various sciences related to computation. A further extension of the 2006-study showing the dynamic trajectories of citations over decades, however, showed that citing audiences fluctuate over time, for example, because of special issues in other disciplines in which cognitive scientists participate (Leydesdorff, Goldstone, & Schank, 2008). In our opinion, patterns of being cited by various audiences should be distinguished from evidence of the inherently interdisciplinary nature of this journal as a central representative of an academic area that regularly spans multiple traditional departments. One can expect the composition of the knowledge bases to be reflected more in the (*citing*) reference patterns within articles than in their reception (Leydesdorff & Probst, 2009).

Our longer term aim is to model the branching and merging of specialties in terms of journal literatures using the tools of both cognitive science and scientometrics, but in this study we focus on the empirical groundwork by providing the bibliometric analysis of *Cognitive Science* in terms of the changes and ramifications of its knowledge base over time. We envisage using the results of this study as heuristics for the future model. To this end, we operationalize the knowledge base of the journal in terms of the journals cited by articles in the journal. The journal has been indexed by Web of Science (WoS) since the fourth edition of 1980, so these journal citations can be retrieved as the subfield of the cited references in the documents. A dedicated routine (BibJourn.exe)[1] enables us to aggregate these cited references into a matrix of (citing) documents versus cited journal names.

Different from our previous studies using aggregated journal-journal citation matrices, the analysis is pursued using documents as units of analysis. The asymmetrical matrix (of documents versus journal names in their cited references) can be imported into SPSS or a network analysis and visualization program such as Pajek for statistical analysis. By limiting downloads to specific publication years of *Cognitive Science*, one can thus dynamically study the changing knowledge bases of the collective production of references by the authors in the journal.

Given that the journal is so purposefully interdisciplinary, this design led first to rather noisy results including 43,952 cited references and 9,911 unique journal names on the basis of the total of 904 (citing) documents. Special issues, for example, disturb the picture. We sought to dampen these effects by using four-year moving aggregates[2] of only the (218) journals that were cited more than 20 times across the entire file (1980-2011). The year 2011 was the last completed year at the time of this research. The choice for an absolute threshold of 20 occurrences was based on visual inspection of the distribution and on the consideration that a percentage threshold might influence the clustering structure in terms of modularity, etc., unevenly. In summary, we have 29 matrices with four-year moving aggregates that we label as “1983” to “2011” by the last completed years. Table 1 provides the descriptive statistics; in order to show the effects of the threshold choice column (d) provides the total numbers of venues involved before setting the threshold (between brackets).

We did not rely on categorizations such as the WoS Subject Categories because these are insufficiently derived formally (Pudovkin & Garfield, 2002: 1113n; Rafols & Leydesdorff, 2009) and are often erroneous (Boyack *et al.*, 2005). *Cognitive Science*, for example, has been attributed in the *Social Science Citation Index* to the journal category of "experimental psychology" (WoS category VX) whereas the journal *Trends in Cognitive Sciences,* covering the same general field, is attributed in the *Science Citation Index* to both "behavioral sciences" (CN) and "neurosciences" (RU), and also to "experimental psychology" in the *Social Science Citation Index.*[3] In sum, the situation is confusing.

[1] BibJourn.exe is freely available at http://www.leydesdorff.net/software/bibjourn/ .
[2] The choice of four-year moving aggregates was made early in the project in order to make sure that not more than 1024 cited journal names would be relevant variables in a single batch given software limitations.

Articles in *Cognitive Science* or *Trends in Cognitive Sciences* may differ in various respects, including their reference patterns, but they share a common subject. In our opinion, the aggregated references provide us with an operationalization of the reference horizons of these communities of authors that can be analyzed both in terms of (cited) authors and journals. The cited authors can be expected to be in flux more than the journal names because the latter are codified and change only as an exception (Bensman & Leydesdorff, 2008). Accordingly, our analyses are in terms of the journals that are cited by articles appearing in *Cognitive Science.*

The use of journals and their aggregated citation relations as indicators of cognitive change has a long history in scientometrics (Price, 1965). The availability of the Journal Citation Reports (in electronic form since 1994) makes such an approach feasible across the file (e.g., Borgman & Rice, 1992; Doreian & Fararo, 1985; Leydesdorff, 1986; Tijssen *et al.*, 1987). More recently, Rafols *et al.* (2010) have proposed overlay maps based on WoS Subject Categories; Leydesdorff and Rafols (2012) followed up with overlays that enable the user to map document sets in terms of journal relations. In an overview of the various mappings, Klavans and Boyack (2009) concluded that a consensus had emerged regarding the structure in science maps based on aggregated journal-journal citation behavior; but Boyack and Klavans (2011) also questioned whether journals are the proper units of analysis for studying (inter)disciplinary structures because journals may themselves be interdisciplinary in their composition of contributions. Our focus on the changing knowledge base of *Cognitive Science* as a purposefully interdisciplinary and branching journal—albeit with a psychology background—may enable us also to throw more light on this question.

Unlike the studies based on journal-journal relations, this study is pursued at the level of documents published in a single journal as units of analysis. From a methodological perspective, we wished also to explore whether newly available software for the dynamic visualization and animation of complex contexts would enable us also to visualize the major dimensions in the set using multidimensional scaling dynamically instead of comparative statics on the basis of factor analysis of each time slice (Leydesdorff, in press; Leydesdorff & Schank, 1998; Mogoutov, personal communication, June 2012).

## Methods and materials

The journal *Cognitive Science* was launched in 1977 and entered the *Social Science Citation Index* with its fourth volume in 1980. As noted, our data consists of the complete set of 904 documents published in this journal in the years 1980-2011, downloaded from WoS on April 25, 2012. These documents contain 43,952 cited references. A regular cited reference in WoS is composed of the name and initial of the first author, publication year, journal abbreviation, volume, and page numbers (e.g., "Hertwig R, 1999, J BEHAV DECIS MAKING, V12, P275"). The third subfield of these references contains the abbreviated journal title. We aggregated this subfield and cross-tabled the resulting frequencies with the citing documents.

[3] Other journals with "cognitive science" in their name like *Topics in Cognitive Science* and *Wiley Interdisciplinary Reviews–Cognitive Science* are attributed to category VX: "experimental psychology."

As can be expected, the frequency distribution of 9,911 unique journal names can easily be fitted to a straight line when plotted on log-log coordinates ($r^2 = 0.92$). We selected the 218 unique journal names that occurred more than 20 times in this set based upon visual inspection of this plot (not shown here). However, the abbreviated journal names are not completely standardized. For example, "Q J Exp Psychol" is cited 44 times, but "Q J Exp Psychol-A" is cited 160 times, and "Q J Expt Psychol" 27 times. We did not correct for these co-referential expressions because they were rare in the case of other journals and this intervention would require the standardization of all journal names equally, whereas the purpose of this project is statistical. However, we corrected for different numbers of the same conference proceedings volumes, such as the "5$^{th}$" or the "7$^{th}$ Ann M Cogn Sci" (that is, the proceedings of the "Annual Meeting of the Cognitive Science Society") and the two ways of indicating this same (and crucial!) conference (that is, "Ann M Cogn Sci" and "Ann C Cogn Sci"). Similarly, in the case of the "P INT JOINT C AR" (that is, the "Proceedings of the International Joint Conference of Artificial Intelligence") the sequence numbers were removed. Furthermore, 587 references to an unpublished "Thesis" were deleted because these titles do not come from a single unified area and thus may distort the interpretation of the disciplinary influences on *Cognitive Science*.

We will use the indication "journals" for the source field in the cited references in the remaining of this study although this may sometimes be conference names or book titles. We removed also all single journal names (in the cited references) from the file. (Misspellings are often occurring only once.) This data cleaning left us with 24,105 (54.8%) cited references. Because cited references in specific years can be biased in terms of special issues in that year, we use four-year moving averages by aggregating, for example, 1980, 1981, 1982, and 1983 into a single file that will be labeled below as "1983" (the most recent year). Thus, the time series runs from 1983 to 2011, and includes 2,987 overlapping documents with 4,212 occurrences of abbreviated journal names in a total set of 29 data matrices.[4] The number of cited journals ranges from 57 journal names in the initial year ("1983") to more than 200 (of the 218 in the entire domain) in each group of four years since "2006".

For each (composite) year, the following files were generated: (1) an SPSS systems file with the documents as the cases and the unique journal-name abbreviations in the respective set as variable labels; (2) a co-occurrence matrix; and (3) a cosine-normalized similarity matrix in the Pajek format.[5] The asymmetrical data matrices are factor-analyzed using orthogonal Varimax rotation in SPSS (v.19). The cosine matrices were reduced to values of cosine > 0.2 for the purpose of visualization. (Without such a threshold, virtually all cells might have a value larger than zero and thus an edge would be drawn in the virtualization among all nodes.) Using the algorithm described by Kamada and Kawai (1989) for force-based spring layout, one can then visualize the various groups for the different years. Gephi was used for another visualization (using ForceAtlas2) and for the computation of network characteristics such as modularity, clustering coefficients, density, etc. Pajek (v3) also conveniently allows data to be exported to VOSViewer as another option for visualization (Van Eck & Waltman, 2010).

---

[4] The total number of documents included is 887 because 17 documents in the download (April 2012) were published in 2012.

[5] Pajek is a program for network visualization and analysis. It is freely available at http://pajek.imfm.si/doku.php?id=download . The Pajek format is also used as a currency among programs in this domain.

Leydesdorff and Schank (2008) developed a dynamic version of the network analysis and visualization program *Visone* (available at http://www.leydesdorff.net/visone) that combines stress-minimization within each matrix and across matrices over time using a weighing factor for static and dynamic stress.[6] This tool was used to capture the results into a streamed shockwave file that can be found at http://www.leydesdorff.net/cognsci/cs.htm . Given that the referencing environment was volatile, we used four consecutive points—each representing agglomerations of four years of data—to achieve further smoothing of the results over time (Gansner *et al.*, 2005; Bauer & Schank, 2008). The dynamic stress adds to the static stress: Kruskall's (1964) aggregated stress for the complete animation of 29 matrices was equal to 0.35.

In addition to the animation, we also explored the development of citations over time by using a recently developed program, CorText, available at http://manager.cortext.net/ . CorText allows for a layout of the dynamics in terms of tubes in an alluvial model (cf. Rosvall & Bergstrom, 2010)[7] that represent components across the cosine-normalized matrices. In our case, we used the 218 most frequently used abbreviated journal titles as input statistics to the analysis but without further pre-processing this data. Precisely as in the other analyses, the cosine was truncated at cosine > 0.2 and the Louvain algorithm (Blondel *et al.*, 2008) used for community detection and modularization. However, the cited references are mapped into the five chunks that the program selects as optimal and not on the basis of our four-year aggregates.

## Results

### *a. Descriptive statistics*

In the presentation below we focus on reporting main trends that are relevant to our study. First, Table 1 provides various descriptive statistics.

---

[6] Pajek projects can also be read by PajekToSVGAnim.exe for the visualization. Given the size of our data, the svg-files become huge (> 60 Mbyte) and svg-files cannot be read by all browsers. Thus, we decided not to use this option.

[7] Rosvall & Bergstrom's (2010) online program for alluvial maps is freely available at http://www.mapequation.org/alluvialgenerator/index.html . However, these authors use a very different similarity criterion and clustering algorithm ("random walks").

| Years (a) | N of documents (b) | Cited journals N > 20 (c) | (N) (d) | Edges (cosine > 0.2) (e) | N of communities (f) | Modularity (g) | Clustering Coefficient (h) | Density (i) |
|---|---|---|---|---|---|---|---|---|
| 1980-1983 | 53 | 57 | (147) | 720 | 4 | 0.267 | 0.537 | 0.226 |
| 1981-1984 | 53 | 63 | (164) | 1050 | 7 | 0.251 | 0.555 | 0.269 |
| 1982-1985 | 56 | 76 | (190) | 1336 | 5 | 0.276 | 0.522 | 0.234 |
| 1983-1986 | 62 | 96 | (266) | 1692 | 5 | 0.321 | 0.489 | 0.186 |
| 1984-1987 | 69 | 99 | (293) | 1798 | 4 | 0.288 | 0.483 | 0.185 |
| 1985-1988 | 72 | 105 | (302) | 1804 | 6 | 0.327 | 0.484 | 0.165 |
| 1986-1989 | 75 | 109 | (311) | 1934 | 6 | 0.303 | 0.474 | 0.164 |
| 1987-1990 | 79 | 110 | (282) | 1922 | 5 | 0.306 | 0.479 | 0.160 |
| 1988-1991 | 77 | 114 | (303) | 2138 | 6 | 0.305 | 0.531 | 0.166 |
| 1989-1992 | 75 | 120 | (304) | 2276 | 5 | 0.320 | 0.520 | 0.159 |
| 1990-1993 | 77 | 128 | (342) | 2542 | 4 | 0.341 | 0.519 | 0.156 |
| 1991-1994 | 73 | 126 | (338) | 2492 | 5 | 0.304 | 0.518 | 0.158 |
| 1992-1995 | 71 | 125 | (339) | 2500 | 6 | 0.312 | 0.516 | 0.161 |
| 1993-1996 | 69 | 126 | (375) | 2530 | 5 | 0.329 | 0.519 | 0.161 |
| 1994-1997 | 62 | 134 | (353) | 2706 | 5 | 0.335 | 0.537 | 0.152 |
| 1995-1998 | 60 | 140 | (385) | 3182 | 6 | 0.329 | 0.525 | 0.164 |
| 1996-1999 | 65 | 150 | (414) | 4160 | 5 | 0.321 | 0.534 | 0.186 |
| 1997-2000 | 70 | 163 | (473) | 4370 | 5 | 0.341 | 0.518 | 0.165 |
| 1998-2001 | 82 | 181 | (597) | 5316 | 7 | 0.330 | 0.505 | 0.163 |
| 1999-2002 | 96 | 189 | (634) | 5248 | 6 | 0.315 | 0.481 | 0.148 |
| 2000-2003 | 115 | 188 | (655) | 4632 | 5 | 0.346 | 0.465 | 0.132 |
| 2001-2004 | 139 | 195 | (676) | 4560 | 5 | 0.377 | 0.446 | 0.121 |
| 2002-2005 | 151 | 197 | (700) | 4244 | 6 | 0.371 | 0.422 | 0.110 |
| 2003-2006 | 164 | 204 | (782) | 3880 | 6 | 0.385 | 0.408 | 0.094 |
| 2004-2007 | 167 | 201 | (806) | 4170 | 5 | 0.395 | 0.442 | 0.104 |
| 2005-2008 | 178 | 205 | (840) | 3998 | 5 | 0.386 | 0.429 | 0.096 |
| 2006-2009 | 203 | 205 | (910) | 3876 | 7 | 0.372 | 0.413 | 0.093 |
| 2007-2010 | 229 | 203 | (976) | 4136 | 7 | 0.327 | 0.415 | 0.101 |
| 2008-2011 | 245 | 203 | (1046) | 4130 | 5 | 0.356 | 0.432 | 0.101 |
| Sum: | 2987 | 4212 | (14203) | 89.342 | Avg. = 5.448 (±0.870) | 0.329 (±0.036) | 0.487 (±0.044) | 0.154 (±0.043) |

**Table 1**: Descriptive statistics of the development of *Cognitive Science* and its knowledge bases during the period 1980-2011

Figure 1 shows that during the period 1980-2011, the journal *Cognitive Science* experienced significant growth in terms of numbers of published articles. The number of source documents has increased precipitously since 2000 because of an editorial decision to allow for brief reports to be published (Goldstone, 2000). Brief reports appeared increasingly over the years since they were approved. These communications tend to have fewer references than regular articles. This transition seems also to have had an initial effect on the density of the journal citation relations, but this effect faded after 2006.

Furthermore, the number of issues of *Cognitive* Science has increased over the years. From 1977-2000, four issues of the journal were published each year. From 2001-2007, six issues were published, and from 2008 up to 2012, eight issues appeared per year. The trends in the various columns of Table 1 may also be partially a result of upward trends in the citation frequencies across the file because of the increasing numbers of references per document (Althouse *et al.*, 2009).

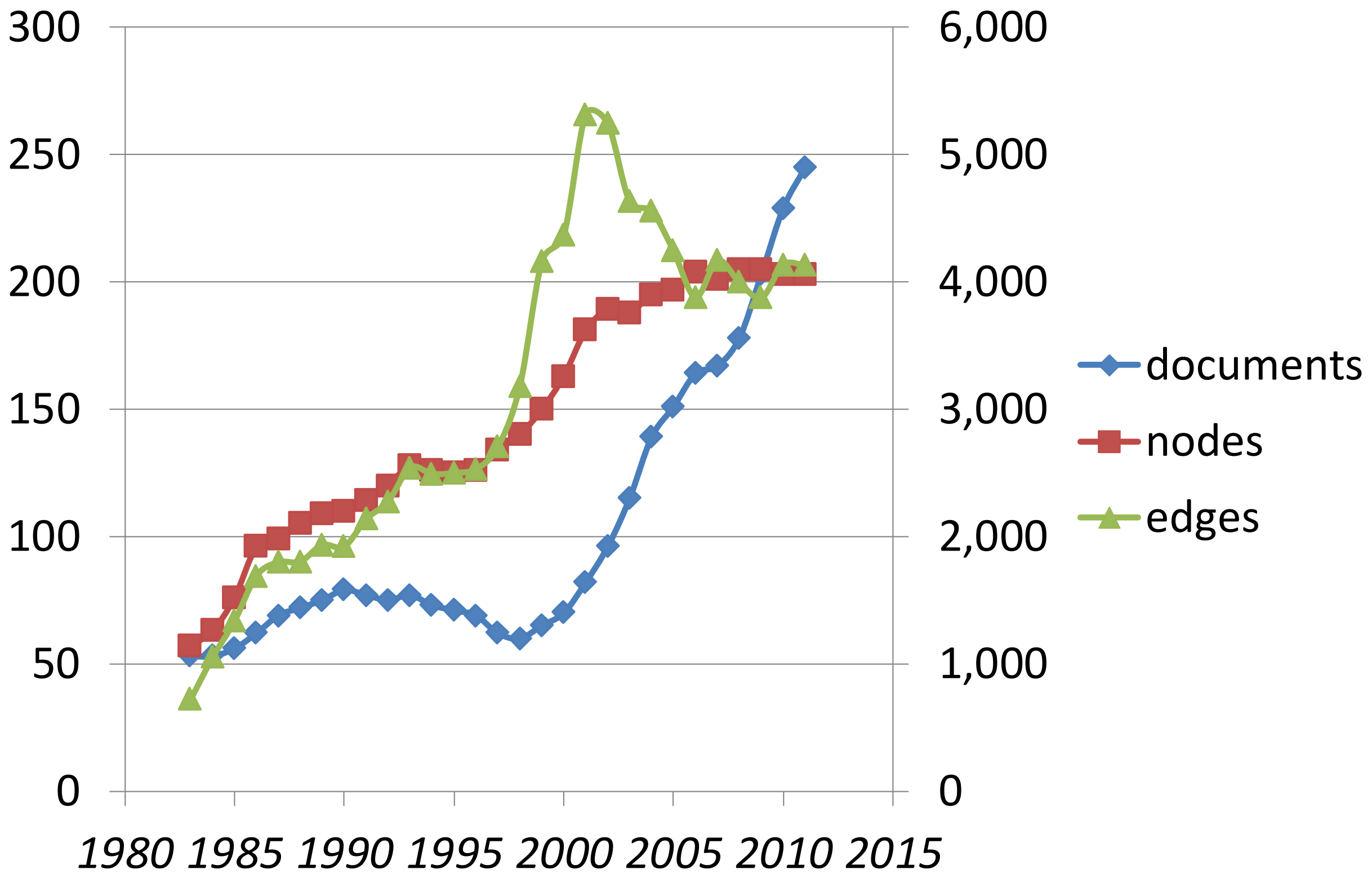

**Figure 1:** numbers of documents (left axis) and the numbers of cited journal names (nodes) and their relations (edges) at cosine > 0.2.

Figure 2 shows that the number of communities detected by using Blondel *et al.*'s (2008) algorithm (in Gephi) fluctuates around 5.5. We decided on this basis to compare five-factor solutions of the matrix in the next section. The density and clustering coefficients decrease with the expansion of the matrix, whereas the modularity increases. Note that these results are based on matrices which are normalized using a cosine transformation. The cosine normalizes differences in size between zero and one using the angle between the distributions (vectors) of cited journals in the citing documents (Ahlgren *et al.*, 2003).

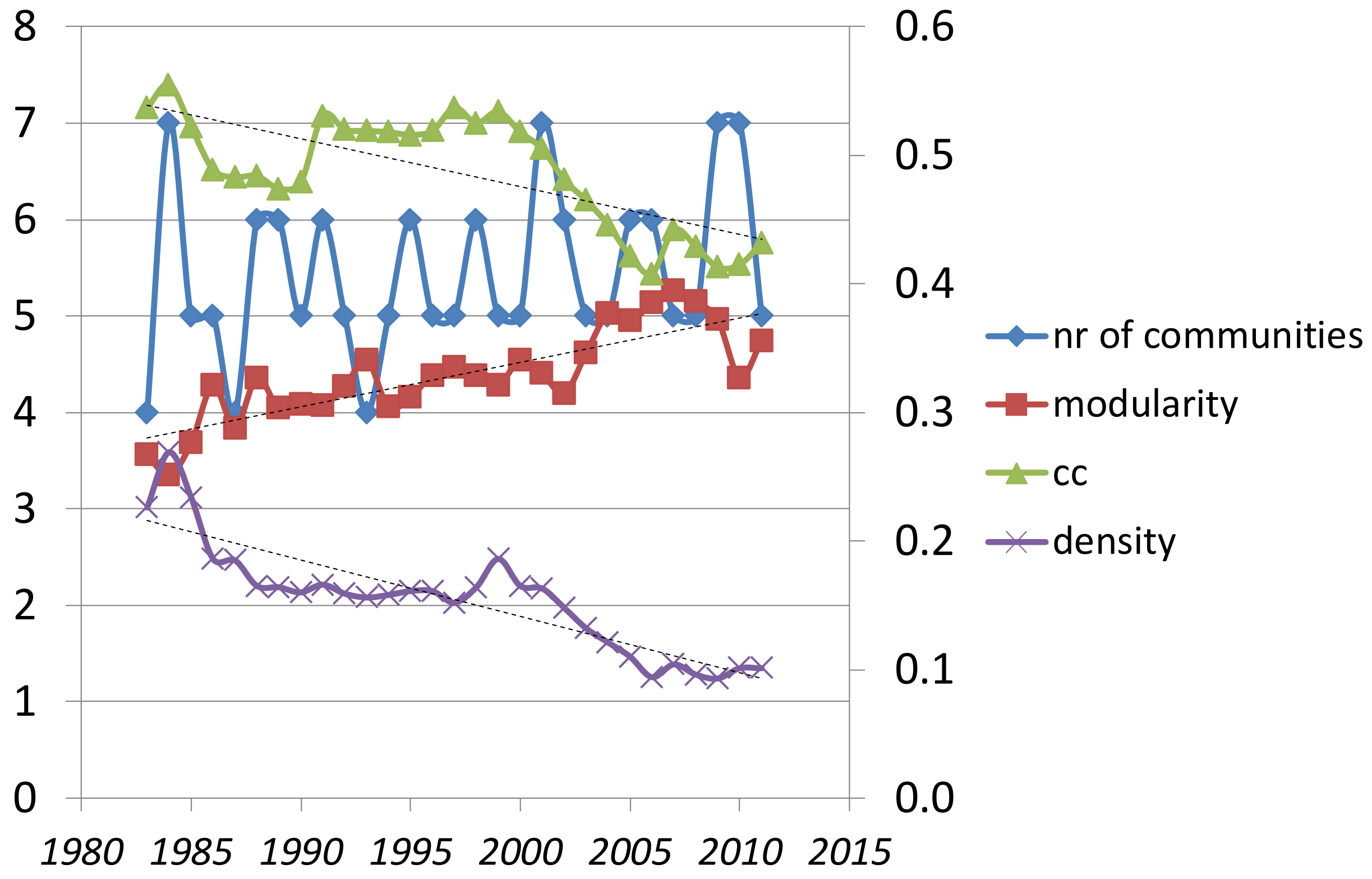

**Figure 2**: The evolution of four network characteristics during 1983-2011: the number of communities (left axis; Blondel *et al*., 2008), modularity, clustering coefficient, and density (right axis).

Figure 2 shows a somewhat increasing modularity over time. This leads to the question of whether one would not expect modularity to increase as a field matures: do articles shift from citing a "common core" of cognitive science to citing articles in specialized subfields which interact decreasingly among themselves (in this environment)? As this journal representing the field increases in volume, it may also become increasingly hard for authors to remain well versed in all aspects of the field. One would expect the journal to draw increasingly on different subfields. The relative constancy in the number of detected communities, however, suggests that the increasing modularity would not be an artifact of larger field sizes being able to support more sub-communities. Instead, it may indicate that the component disciplines to which authors in *Cognitive Science* refer become increasingly more specialized. This conclusion is also supported by the decreased clustering coefficient over time. This coefficient reflects the tendency of two journals cited by articles in a third journal to cite each other.

In summary, a coherent pattern emerges from the indicators presented in Figure 2. As the field increases in size, the density of connectivity among the cited journals decreases—journals that are cited by other journals become less likely to cite each other—and there is increasing compartmentalization of the broader intellectual environment surrounding *Cognitive Science* into fields. All of these trends are consistent with authors having a limited capacity for reading articles; as the total field increases in size, they cope with this limited capacity by narrowing the scope of their reading/citing to fewer (sub)fields within cognitive science.

*b. Static and dynamic visualizations*

During the period of our project, new visualization and analysis tools have become available, such as the smooth integration of VOSViewer (Van Eck & Waltman, 2010) and Blondel *et al.*'s (2008) algorithm for community detection in Pajek v3 in July 2012. A plethora of visualizations is thus possible with differences in the number of communities and visualization techniques. Figure 3, for example, shows the most recent (that is, 2008-2011) map of six communities detected by the same algorithm as above and using VOSViewer for the visualization. (The colored version of this figure can be webstarted at http://www.vosviewer.com/vosviewer.php?map=http://www.leydesdorff.net/cognsci/figure3.txt&view=3&zoom_level=1 .)

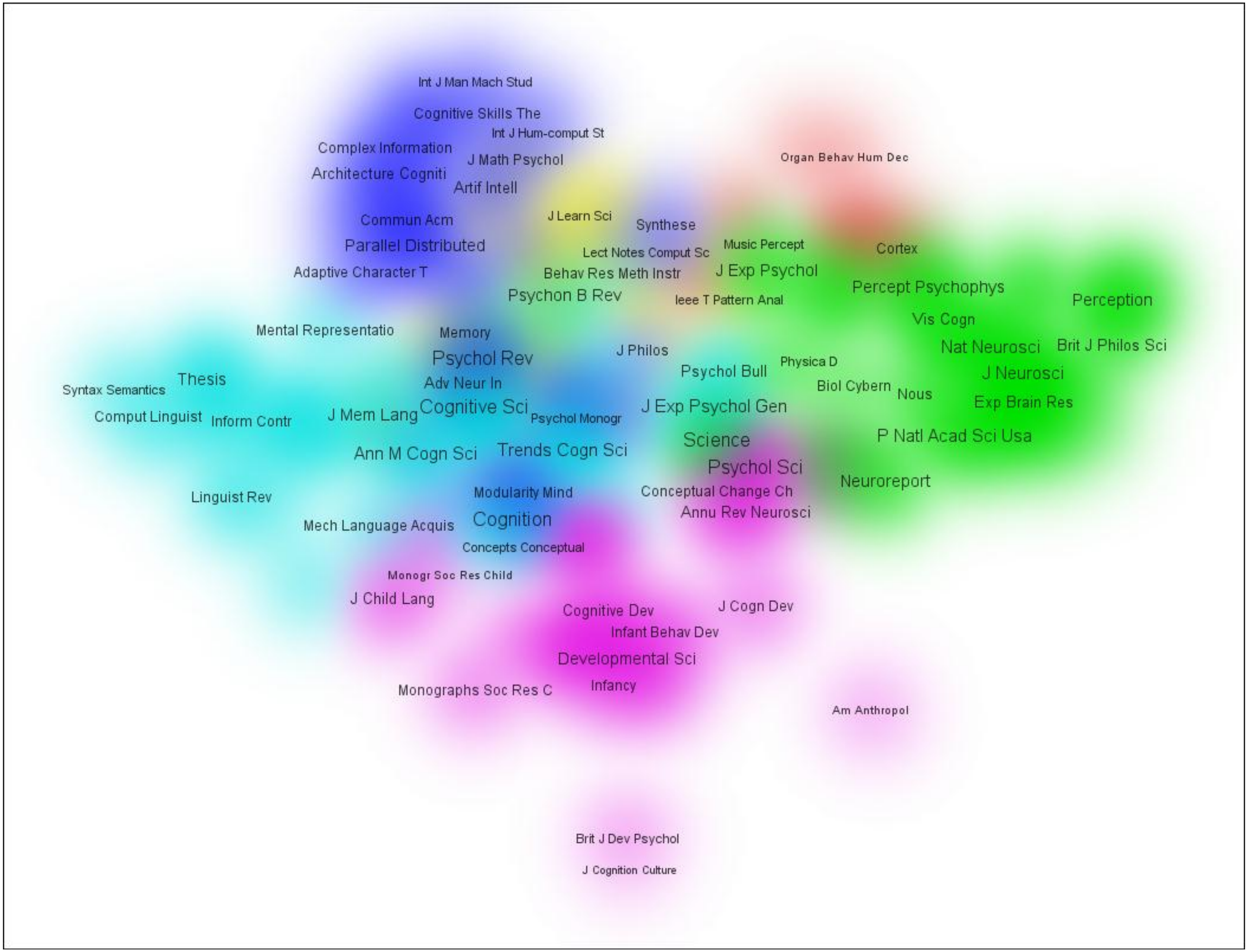

**Figure 3**: Six components using Blondel *et al.*'s (2008) modularity algorithm in Pajek on the basis of 203 journals cited in 245 documents published by *Cognitive Science* during 2008-2011; VOSViewer used for visualization. This map can be accessed interactively at http://www.vosviewer.com/vosviewer.php?map=http://www.leydesdorff.net/cognsci/figure3.txt&view=3&zoom_level=1 .

Figure 3 shows the grouping of neuroscience journals on the right side (in green) and language on the left side (light blue). Journals in the computer sciences and artificial intelligence are positioned at the top (dark blue), and psychology journals about cognitive development are shown at the bottom (pink). Two further groups are distinguished: one in yellow focusing on cognitive instruction and education, and one in orange with journals in experimental psychology and decision theory. Our target journal *Cognitive Science* is assigned to the language-oriented group by this algorithm (Blondel *et al.*, 2008), while positioned in the central area of the map.

The VOSViewer visualization has the technical advantage of presenting a heat map, and the labels of the most-connected journals in the network are foregrounded. Furthermore, the visualization is based on an MDS-like algorithm. One of us has argued elsewhere why this combination of choices might be optimal for representing cosine-based maps in terms of a vector space (Leydesdorff, in press; Leydesdorff & Rafols, 2012). The groupings, however, remain sensitive to parameter choices because the environment is relatively fuzzy and volatile.

The animation using *visone* at http://www.leydesdorff.net/cognsci/cs.htm deliberately counteracts this volatility by minimizing the stress value over time.  Linguistics, for example, dominates the upper-right corner of the animation, whereas psychology is at the left, but moves somewhat more to the bottom in more recent years. The journal *Cognitive Science* is indicated as a red node positioned in the central region among the disciplines in all years. The disadvantage of this representation, however, is the difficulty in drawing delineations between fields. The impression of field compartmentalization changes over the years. We could have further refined the animation by coloring the nodes in terms of the solutions for different years, but as we shall see below (using factor analysis), both the vectors and the eigenvectors change relative positions during the analyzed period. For example, the "neurosciences" become more important in later years (the 2000s).

Before turning to factor analysis to study these evolving structures in a comparative-static design, however, let us focus on another tool "CorText" (at http://manager.cortext.net) that seemed highly appropriate for our purpose and was introduced to both of us by Andrei Mogoutov, its developer, in June 2012 during a visit to Amsterdam.

### *c. Flow models of merging and splitting using CorText*

Independently of us, but using the same literature about the dynamic visualization of multivariate data, the authors of "CorText" (available upon registration at http://manager.cortext.net) had reasoned along very similar lines, but in addition to cosine-based maps, this program allows for dynamic analysis of field components using (so-called "Sankey") flow-diagrams of the citation relations among the journals in the set. Rosvall and Bergstrom (2010) first introduced these maps into bibliometrics as "alluvial diagrams" (see at http://www.mapequation.org/alluvialgenerator/index.html ). However, Rosvall and Bergstrom (2010) based their visualization on other statistics such as random walks and information theory.

Similar to our design, CorText allows for the cosine to be used as the similarity measure, and the Louvain-algorithm (Blondel *et al.*, 2008) for the decomposition. Furthermore, data can be imported from WOS, and the subfield of journals in the "cited references" of these documents can be chosen as nodes in the network visualizations. The number of nodes and edges can be specified.

The program self-optimizes the number of time slices to consider for determining important changes. We chose the default of five time slots, and we also left all other choices to the suggestions made by the program and in consultation with its main author (Andrei Mogoutov). The threshold for the cosine was set at 0.2 by the program, similar to our choice, and we asked for the 218 most-cited journals included in our set as the nodes, with 89,342 edges as specified in Table 1 above.

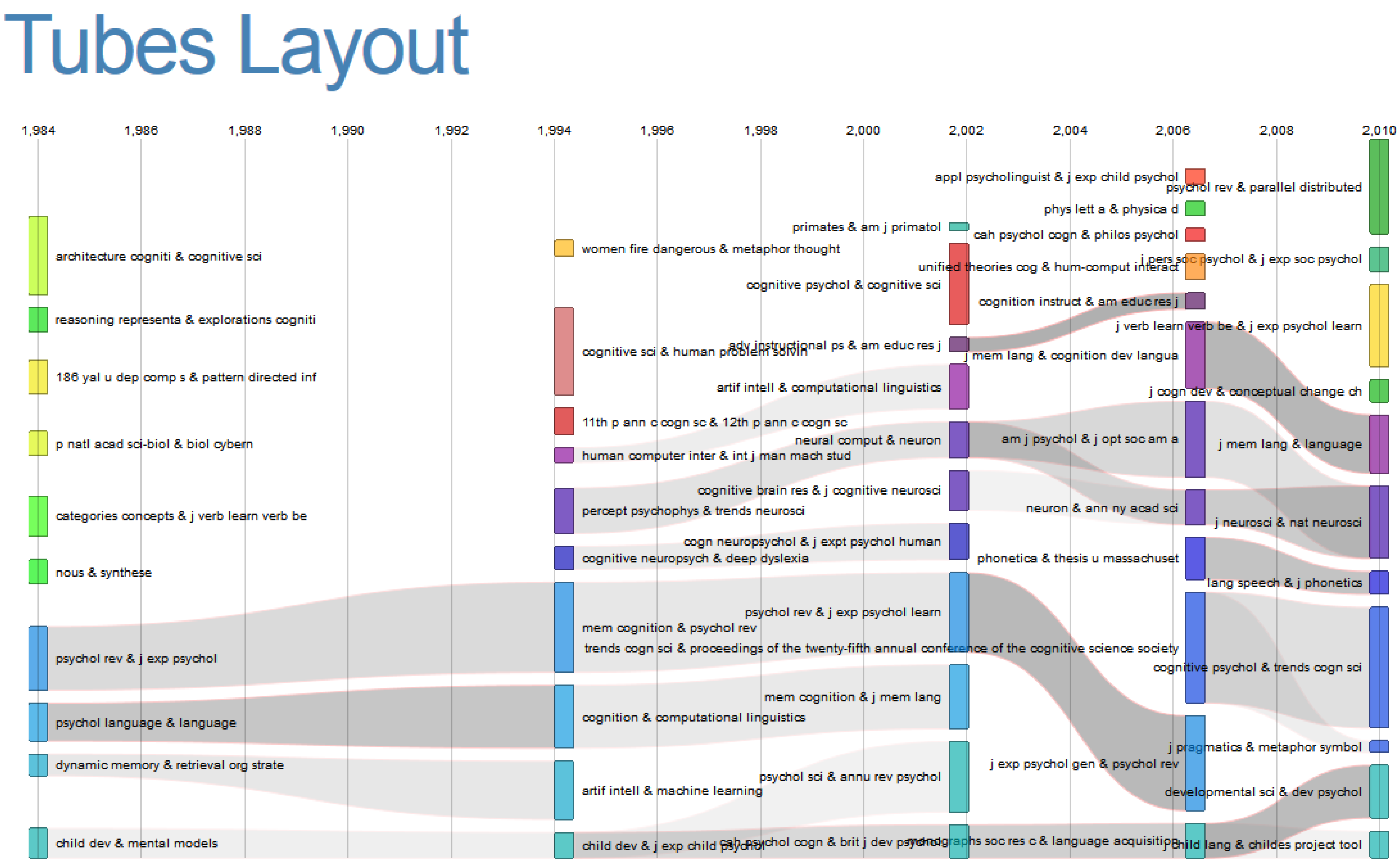

**Figure 4**: Tubes Layout of CorText using the complete set of 887 documents and 43,284 cited references.

The results, shown in Figure 4, are somewhat different from the results of our earlier analysis. As before, these results are sensitive to specific parameter values, but the trends tend to remain across parametric variation. First, the complexity in the field increases—according to this routine—from four communities in the period 1984-1994 to six in 1994-1996, and seven during 1997-2002. After 2002, the number of communities decreases to five or six. The reorganizations in the flow diagram shortly before 2002 and then after 2006 correspond with new managing editors in office since 2001 and 2006, respectively. We note this coincidence, but do not have sufficient reason to believe that there is causal relationship (Zsindely *et al.*, 1982)..

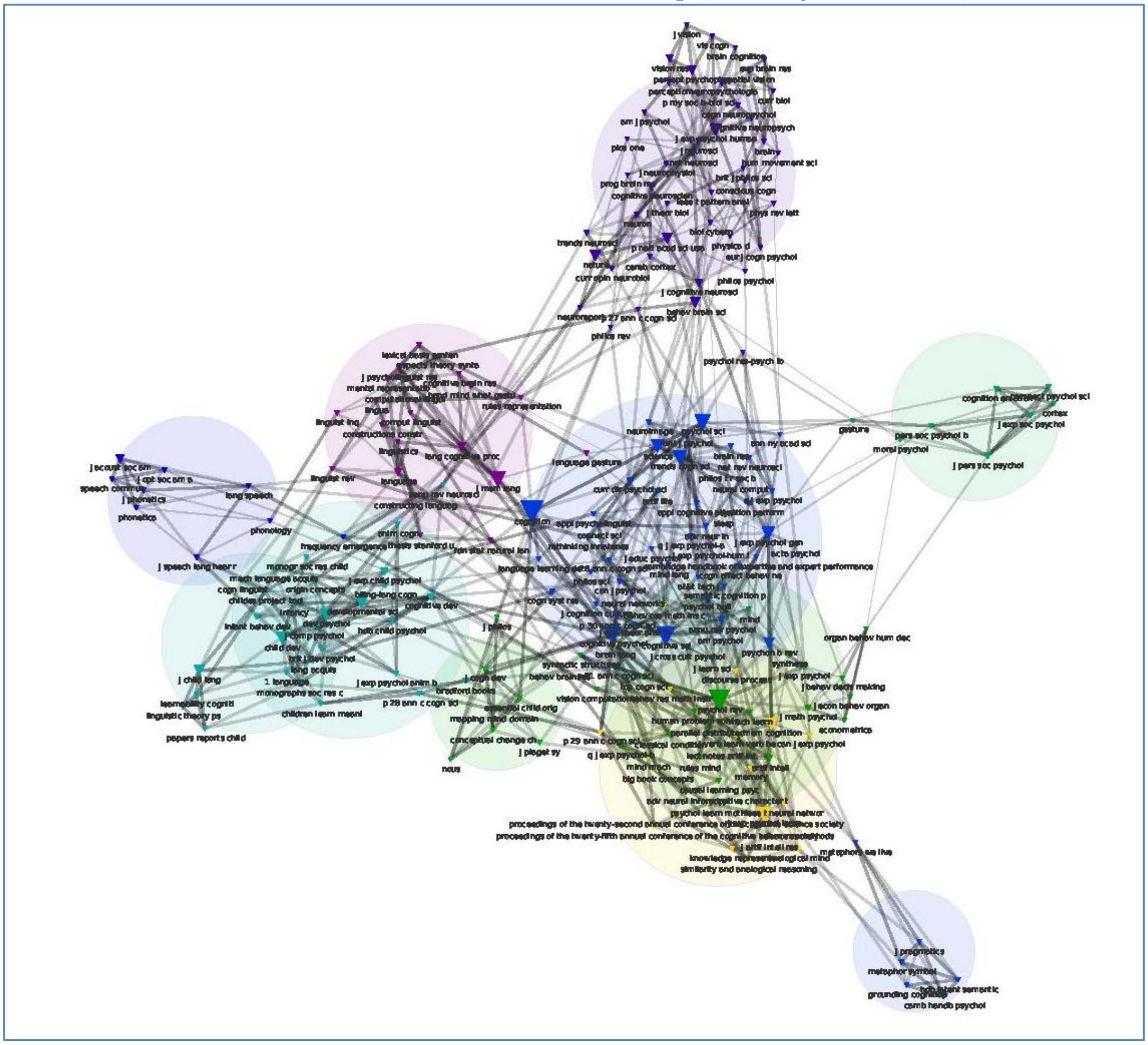

**Figure 5**: cosine-normalized map of 203 cited journals in the most recent slide (2009-2011) based on CorText. (A larger, interactive version is available at http://www.leydesdorff.net/cognsci/figure5.htm.)

In addition to the flow maps, CorText also provides maps for the different time periods. Figure 5 shows the cosine-normalized map of 203 cited journals included during the most recent period. (The legends cannot be read in this printed version, but an interactive version is provided at http://www.leydesdorff.net/cognsci/figure5.htm.) Although this map is based on the same threshold (cosine > 0.2) and the same community-finding algorithm, the number of communities is now indicated as much larger, and corresponds to the eleven groups in the last bar (2010) of Figure 4. However, we were not able to clarify these somewhat confusing differences.

Whereas the number of flows in Figure 4 fluctuates more or less in agreement with the numbers of communities in Figure 2, the single journal names associated with each flow are difficult to designate in disciplinary terms. The relations between Figures 4 and 5, furthermore, are insufficiently clear to warrant an interpretation. It appears that splittings within communities may be more common than fusions of two communities. Figure 4 shows four cases of splittings compared with 2 fusions. This result is consistent with the increasing modularity depicted in Figure 2, but given the small total number of splitting and fusion events, it is important not to over-interpret these results in isolation.

Let us caution that CorText is still in its developmental phase. However, the current implementation did not allow us to further clarify our research question because too many questions can be raised about the origins of the differences between the resulting visualizations and the maps that we were able to generate using the same similarity criteria and clustering algorithm.

**Factor analysis**

Because of the inconclusiveness of the above analyses based on different visualizations, we return in this section to a more traditional approach using the documents as units of analysis and the cited journal names as the attributed variables. Each matrix is then based on the number of documents in column (b) of Table 1, and the number of variables as specified in column (c) of this same table. Our data is composed of three decades (1980-2011), but in terms of 4-year aggregated time periods. We focus on the aggregates of (*i*) 1985-1988 as probably informative about the early period of the journal and the intentional construction of the interdiscipline; (*ii*) the period 1995-1998 as representative for the period of relative stabilization and interdisciplinary development; and (*iii*) the period 2005-2008 including the turn to "neuro" in psychology. The input matrices contain whole number counts: in other words, if a journal is cited twice by the same document (but with a reference to another cited document) then the cell value is two (etc.). From Table 1 (column c), one can see that the first period included 105 journals, the second 149, and the third 205.

| | *1985-1988* | *1995-1998* | *2005-2008* |
|---|---|---|---|
| ***% variance explained*** | 27.8 | 28.5 | 18.6 |
| ***Factor 1*** | Exp. psychology *Q J Exp Psychol* | Language *J Psycholinguist Res* | Neurosciences *Neuron* |
| ***Factor 2*** | **Language** *Language Acquisition* | Biology *P Natl Acad Sci USA* | Perception, sensation, etc. *Perception* |
| ***Factor 3*** | Cognitive psychology *J Exp Psychol Learn* | Learning and development *Child Dev* | Learning and development *Dev Psychol* |
| ***Factor 4*** | Linguistics *J Mem Lang* | Cognitive psychology *Psychol Learn Motiv* | **Cognitive psychology** *Psychol Learn Motiv* |
| ***Factor 5*** | Philosophy & AI *J Philos* | **Computation & philosophy** *Human Computer Inter* | Psychology / AI / language *Psychol Rev* |

**Table 2**: Results of the factor analysis; 5 factors extracted; Varimax rotated. Journals with highest factor loadings indicated. Factors with the major factor loadings for the journal *Cognitive Science* are boldfaced.

The screeplots for all three periods suggest the extraction of four to five factors. Given the modularity of approximately five to six groups in these years, we use the 5-factor solution for the comparison in Table 2. The factor designation is ours, and therefore we provide also the journal names with the highest factor-loadings on the corresponding factor (using the standard abbreviations of WOS for cited references). The development is very turbulent, but this can be illustrated in considerable detail using this table. We boldfaced in Table 2 the factor designations on which the journal *Cognitive Science* had the highest loading in this period. However, the loadings for this journal show considerable factorial complexity in all periods. Van den Besselaar and Heimeriks (2001) consider this as an indicator of interdisciplinarity in citation patterns among journals (cf. Van den Besselaar & Leydesdorff, 1996, at p. 428).

### *a. The generation of the discipline during the 1980s*

During the first period, *Cognitive Science* has a low loading ($r$ = 0.111) on the second factor that is otherwise indicative of experimental approaches to language. On the first factor, designated by us as "experimental psychology," and the fourth factor indicating formal linguistics, however, the journal has negative loadings. Thus, the relation to these two "mother" disciplines is expressed as differences in aggregated citation behavior. The journal belongs to a group of journals with specific reference horizons.

As could be expected, "philosophy" also plays a role during this period in relation to the journals that will belong in the next period to "artificial intelligence." For example, *International Journal for Man-Machine Studies* has its highest factor loading (0.547) on this fifth dimension, as do *Artificial Intelligence* (0.229) and *AI Magazine* (0.530). Van den Besselaar and Leydesdorff (1996) indicated 1986-1988 as the transition period towards a separate cluster of journals representing AI. It was not, however, found meaningful to consider *Cognitive Science* as a separate grouping during this same period.

*b. interdisciplinary development during the 1990s*

In the second decade, *Cognitive Science* further developed into an interdisciplinary platform where cognitive psychologists (Factor 4) met authors focusing on language, development, computation, and philosophy, and the various knowledge bases were integrated into reference patterns. The journal has positive factor loadings[10] on four of the five factors distinguished, but a negative factor loading on the second factor; this second factor indicates a common variance among natural science and biology journals. The development of the journal during this period thus accords with its ambition to be a leading journal at the interfaces of these disciplines, notably in terms of computational approaches to problems of (child) development, language, and cognition. The biological dimension is present, but at this time is still referenced as specific and different.

*c. integration into cognitive psychology during the 2000s*

This interdisciplinary orientation changes during the third period. The biology factor now becomes the most pronounced one, followed by a "perception" factor on which journals focusing on perceptions, sensation, vision, etc., load. *Cognitive Science* loads negatively on this latter factor; its main factor loading (.210) is on the factor that can be designated as cognitive (as different from developmental) psychology. In the meantime, cognitive psychology has itself become a more coherent specialty, and *Cognitive Science* has been integrated into this specialty within psychology more than before. Has the interdisciplinary ambition been incorporated into psychology?

A second, but minor factor loading (.120) during this period is on Factor 5 which groups journals on the margins of artificial intelligence, linguistics, and experimental psychology. The shift towards "neuro" in psychology, however, has affected the position of *Cognitive Science* in these environments, and as noted above, authorship has also become more oriented toward psychology. The interdisciplinary orientation seems to have waned, and become secondary to the affiliation with psychology. One can also observe that psychology has become more receptive to interdisciplinary developments (including "neuro") or, in other words, the innovation and mission of *Cognitive Science* has been effective within the ongoing changes in the mother discipline.

## Discussion

It is interesting to compare the development of *Cognitive Science* with that of the journal *Artificial Intelligence* which Van den Besselaar and Leydesdorff (1996) have studied previously. Both journals can be considered as interdisciplinary projects in approximately the same domain when they were launched in 1979 and 1970, respectively, and in each other's environments at the interfaces between computer science, psychology, and linguistics. In this previous research, we found that between 1986 and 1988 AI became a specialty after a confluence between the citation patterns of *Artificial Intelligence¸ AI Magazine,* and *IEEE Expert* (renamed *IEEE Intelligent Systems & Their Applications* in 1997). Ever since, this cluster of currently 10+ journals has grown into a recognizable specialty structure with its own reproduction mechanisms such as departments, curricula, and conferences. One can surmise that AI lost its interdisciplinarity as a specialty in the late 1980s and early 1990s, or that "interdisciplinarity" should be defined differently at the specialty level and at the journal level (Wagner *et al.*, 2011).

[10] For technical reasons, Factor 5 has an opposite sign, but the journal *Cognitive Science* follows with the same sign.

During the 1990s, *Cognitive Science* remained a specialized journal that continued to explore new options for interdisciplinarity at the relevant interfaces, but from a starting position in psychology more than computer science, philosophy, linguistics, or education. The sister journal *Trends in Cognitive Sciences* was launched in 1997, and *Topics in Cognitive Science* in 2009. These journals and other similar ones in their environment have, however, not managed to break away from their disciplinary base in psychology.

During the 2000s, furthermore, institutional incentives have been influenced by university rankings and consequent evaluations in terms of disciplinary frameworks, and interdisciplinary ventures have become more risky (Rafols *et al.*, 2012). In this context, these journals had an option to become part of a growing set of journals within psychology that focus on cognition, neuroscience, development, and language acquisition. The momentum of innovation at the interstices between disciplines may have lost its attraction in terms of potential audiences (and hence citations).

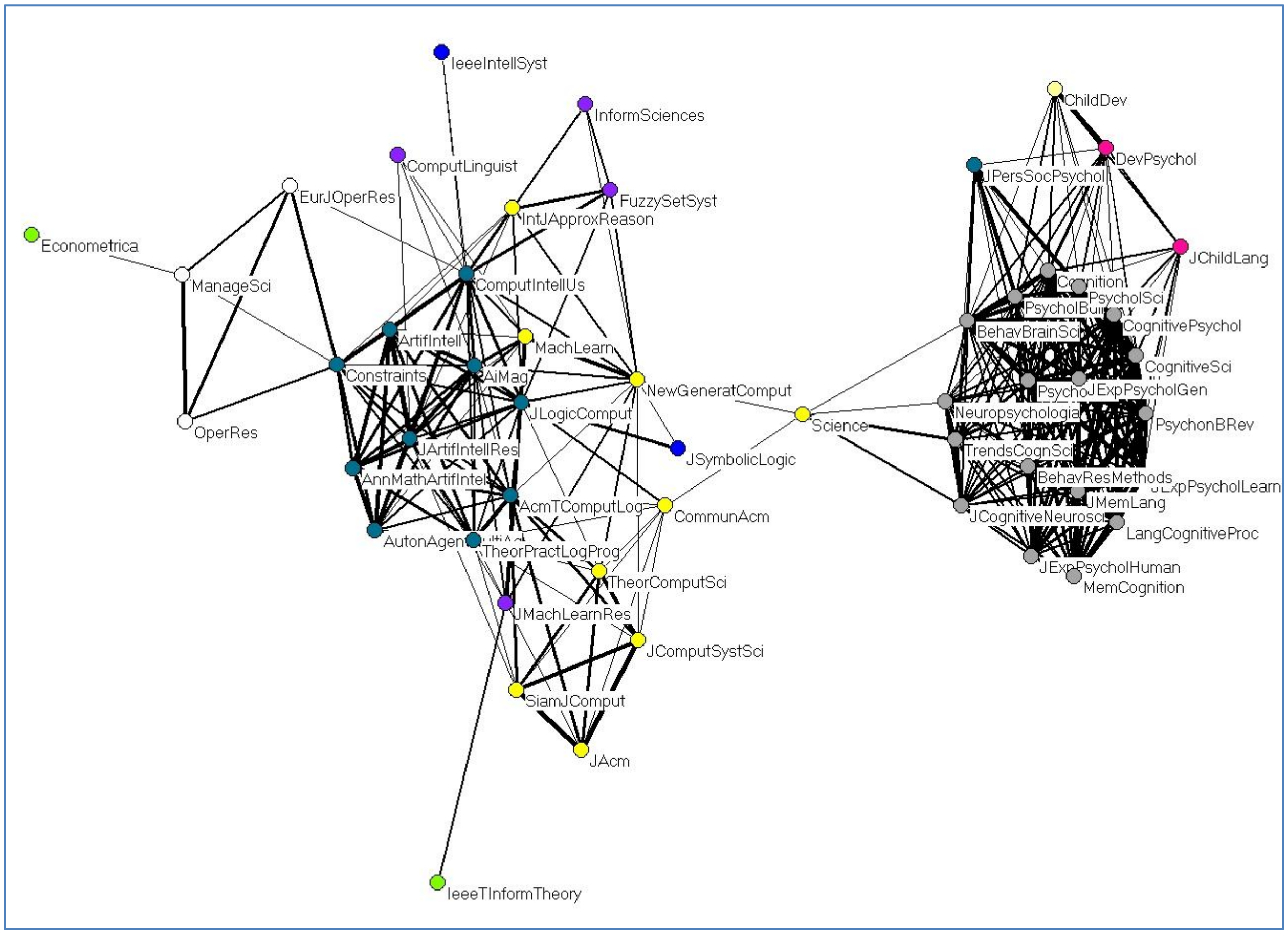

**Figure 6**: Citation network among 30 journals most often cited in *Artificial Intelligence* in 2011 and 22 journals most cited in *Cognitive Science*, respectively, to the extent of more than 1% of each journal's total citations; cosine > 0.2; *k*-core algorithm used for the coloring.

Figure 6 shows the (cosine-normalized) aggregated citation relations between the 32 most cited journals in *Cognitive Science* in 2011 and the 20 most cited journals in *Artificial Intelligence* for the same year.[11] Using this journal mapping technique, the two domains are now completely separate except that *Science* is cited in both contexts as a multidisciplinary journal. *Cognitive Science* is firmly embedded in a set of psychology journals, whereas *Artificial Intelligence* is part of a domain of journals focusing on artificial intelligence as a specialty structure.

In summary, the earlier conclusion of Van den Besselaar and Leydesdorff (1996, at p. 428; cf. Van den Besselaar, in preparation) that the interdisciplinary citation environment of *Cognitive Science* could not be stabilized can be confirmed by this study at the document level. These authors indicated the interfactorial complexity—that is, the loading on more than one factor—as typical for interdisciplinarity and transition (Van den Besselaar & Heimeriks, 2001). Whereas AI made the transition towards establishing a specialty structure, *Cognitive Science*'s intellectual niche has settled within the domain of psychology. The field of psychology has undergone important changes in terms of the (inter)disciplinary horizons of referencing during the last two decades.

## Conclusions

The analyses reveal a number of interesting trends with respect to cognitive science in particular, and the study of interdisciplinarity more generally.

### *a. Cognitive science*

Specific to cognitive science, several interesting trends are apparent in the clustering, alluvial, and factor analyses. First, neuroscience has become more central to cognitive science in the most recent decade. Second, issues pertaining to development have become more unified and can now be distinguished more clearly from other fields pertaining to the reference horizons of cognitive scientists. Third, there has been an increasing separation between cognitive science and philosophy, and to a lesser extent artificial intelligence. Fourth, there is a definite risk/tendency (as noted by Gentner, 2010; Thagard, 2005, 2009) for cognitive science to be dominated by psychology departments. Although they have been distinguished by some of our analyses, neuroscience, perception, development, cognitive psychology, and experimental psychology are all activities of psychology departments. Some of the eclecticism of earlier cognitive science, with strong contributions from linguistics, philosophy, and AI, seems to have been pushed aside by new developments at the disciplinary level and the consequent reformulations of the missions of psychology departments.

### *b. The inherent fragility of interdisciplinarity*

Psychology has always been the dominant field within cognitive science. In earlier years, computer science, linguistics, and philosophy played crucial and major roles. However, if cognitive science began with a slight dominance by psychology, and psychology has strong within-field relations, then it might (from the perspective of hindsight) have been predictable, perhaps inevitable, that psychology would become even more dominant within cognitive science, at the same time that new areas from within psychology become more robustly represented (Cacioppo, 2007). The original core of cognitive science from within psychology can be interpreted imperfectly as "cognitive psychology." Since the early days of *Cognitive Science,* other fields from psychology have become increasingly brought into the fold of the journal: development, neuroscience, sensation, and social psychology.

[11] The two sets were generated using a threshold of 1% of the total citations excluding journal self-citations.

*Cognitive Science* presents an interesting case study in the development of an interdisciplinary ambition over time. One might have posited that an originally interdisciplinary field would become specialized into sub-fields over time, with the sub-fields having increasingly less interconnectivity. We find some evidence for this in the increasing modularity of field components over time (see Figure 2). One might have conjectured also the converse trend, with the interdisciplinary field bridging originally disconnected fields with the result that the fields subsequently become more interconnected.

Neither the modularity nor factor-analytic results provide much evidence for this dynamic. However, a third possibility becomes apparent by considering the larger ecology of other fields in which an interdisciplinary field resides. Imagine that an interdisciplinary field V (cognitive science to be specific) crucially involves a part of another field W (psychology) but also involves to a somewhat lesser extent fields X (computer science), Y (linguistics), and Z (philosophy). Imagine further that W is tightly integrated with robust intra-field connections. In this case, V may change to adopt increasingly more components of W. When combined with a richer-gets-richer dynamic, the result can be that field V becomes increasingly similar to W, and less similar to X, Y and Z.

Virtually all cognitive scientists continue to champion the interdisciplinarity of their research area, and similar calls-to-arms for interdisciplinarity have been made in the related research areas of emotions (Kappas, 2002) and information science (Holland, 2008). From its very inception (Collins, 1977) through to the new millennium (Schuun, Crowley, & Okada, 1998; Von Eckardt, 2001), cognitive scientists have repeatedly made claims for a truly interdisciplinary field of cognitive science. Despite this, the genuine interdisciplinarity of cognitive science is decreasing, not increasing.

A useful analogy for the increasingly disproportionate representation of one field within an interdisciplinary enterprise is provided by Schelling's (1971) classic simulation studies of segregation. Schelling created agents belonging to two classes that are reasonably tolerant of diversity and move only when they find themselves in a clear minority within their neighborhood, following a rule like "If fewer than 30% of my neighbors belong to my group, then I will move." Despite this overall tolerance, the agents still divide themselves into sharply segregated groups after a short time. What is surprising is that this occurs even though no individual in the system is motivated to live in such a highly segregated world. Although hardly a realistic model of migration, the model was influential in contrasting group-level results (i.e. widespread segregation) and individual goals.

Likewise, the gradual takeover of an interdisciplinary field by one of its components may be a nearly inevitable consequence of the broader intellectual ecology in which the field has formed its niche. There is a very real competition between different carvings of the intellectual pie. The fate of an interdisciplinary enterprise such as cognitive science is affected not only by its own internal unity and intellectual justification. It is also influenced by the connectivity of its components to other fields. As a result, in understanding the evolution of scientific fields, an important third dynamic to add to "field splitting" and "field fusing" would be the potential for "assimilation into pre-existing fields."

*c. Comparing and contrasting case studies of interdisciplinarity*

In addition to structures indicating established disciplines, new ventures in the (social) sciences are indicated by new journals that take risks at the margins between disciplines. The successful bridging among (sub)disciplines is a relatively rare event. In previous studies, for example, Leydesdorff and Schank (2008) found that the journal *Nanotechnology,* supported by a similar function of *Science,* played such a catalyzing role at the end of the 1990s in establishing interfaces between established specialties in applied physics and chemistry leading to the formation of nanoscience and nanotechnology in the early 2000s.

In the social sciences, Leydesdorff and Probst (2009) traced the emergence of communication studies during the second half of the 1990s, but more in terms of sets of journals that gradually became more densely networked into a new specialty (cf. Rice *et al.*, 1988; Rogers, 1999). A similar dynamic happened in the domain of AI in the late 1980s. Milojevič and Leydesdorff (in press) most recently pointed to the concentration of a subdiscipline of bibliometrics within the field of library & information science, whereas Leydesdorff and Van den Besselaar (1997) showed how and why the interdisciplinary specialty of Science & Technology Studies (STS) has remained at risk of disintegration (cf. Martin *et al.*, 2012).

During and after a transition into a specialized journal set, institutionalization can be a major driver of new developments. The new specialty develops in terms of its own curricula, Ph.D. programs, conferences, etc. One needs these institutions for academic survival (Rafols *et al.*, 2012), and if institutionalization is not achieved, there may be no other option than a return to the mother discipline and a relabeling of the history of the interdisciplinary venture as a renewal of existing structures. As in the case of business ventures, one can consider these two modes of evolution as "creative destruction" (Schumpeter, 1939) versus "creative agglomeration" (Soete & Ter Weel, 1999) or, in another terminology: "Schumpeter Mark I" and "Schumpeter Mark II" (Freeman & Soete, 1997). One either innovates at the margin and succeeds, or one uses the margin to innovate with a feedback arrow to the existing structures.

What does this mean for the concept of "interdisciplinarity?" In our opinion, "interdisciplinarity" should always be specified with reference to a system under study. A research program can be interdisciplinary; a research institute can bring together scholars from different disciplinary backgrounds; a journal can deliberately aim at crossing disciplinary boundaries; or even a specialty can become more interdisciplinary than usual because of the contributions from scholars with different backgrounds. In the case of communication studies, for example, Leydesdorff and Probst (2009) found that interdisciplinary backgrounds remained reflected in citing behavior—because of the participants' scholarly backgrounds and education—whereas in the cited patterns of these journals the relevant environments no longer distinguished these backgrounds.

A reduction of complexity in the environment to two or perhaps three disciplinary identities may be a condition for making the transition to institutionalization (Leydesdorff, 2011; Leydesdorff & Schank, 2008, at pp. 1816f.). For example, STS, which is composed of contributions from sociology, economics, science policy analysis, and scientometrics, has been pulled apart because the center of the field could not be stabilized beyond local manifestations (such as relatively specialized conferences). Communication Studies has been able to shield itself in terms of strong borderlines between its core literature and, for example, that of the information sciences and sociology. AI has grown into a disciplinary structure in the meantime. *Cognitive Science* may have remained too programmatic in its specification of "interdisciplinarity" as "reaching out" from psychology, so that no firm and unambiguous bridges could be established. Interdisciplinarity is then defined at the journal level and insufficiently at the field level.

Like other institutions, journals are specific organizations in which different types of communication can be brought together and interfaced. Specialty structures develop above the journal level, that is, in terms of sets of journals. A new journal may be able to trigger a transition at this next-order level, as in the case of occupational hygiene during the second half of the 1970s (Leydesdorff, 1986) or nanotechnology in the late 1990s. One is able to follow these developments in terms of common variances in citation patterns that can be designated as latent factors (Leydesdorff *et al.*, 1994). The robustness of the emerging structures can thus be tested.

In accordance with a cybernetic principle, the construction of an identifiable eigenvector in the (latent!) next-order structure is bottom-up, but control tends thereafter to become increasingly top-down. As the new paradigm becomes established it feeds back; and the nature of this feedback determines whether the old structures differentiate internally or a bifurcation takes place. In the case of *Cognitive Science,* unlike AI, such a bifurcation seems not to have been needed and the journal could be absorbed into "cognitive psychology" as its basin of attraction.

**Acknowledgement**
We are grateful to Andrei Mogoutov and Peter van den Besselaar for comments and suggestions, and to Thomson-Reuters for allowing to use their data.